# Josephson coupling in high-$T_c$ superconducting junctions using ultra-thin BaTiO$_3$ barriers

H. Navarro, [1*†] M. Sirena,[1,2] Jeehoon Kim,[3,4,5] N. Haberkorn.[1,2]

[1] Instituto Balseiro, Universidad Nacional de Cuyo and Comisión Nacional de Energía Atómica, Av. Bustillo 9500, 8400 San Carlos de Bariloche, Argentina.

[2] Comisión Nacional de Energía Atómica and Consejo Nacional de Investigaciones Científicas y Técnicas, Centro Atómico Bariloche, Av. Bustillo 9500, 8400 San Carlos de Bariloche, Argentina.

[3] Department of Physics, Pohang University of Science and Technology, Pohang 37673, Republic of Korea.

[4] Advanced Material Science, Pohang University of Science and Technology, Pohang 37673, Republic of Korea.

[5] Max Planck POSTECH Center for Complex Phase Materials, Pohang University of Science and Technology, Pohang 37673, Republic of Korea.

**Corresponding Author**
*E-mail: hnavarro@physics.ucsd.edu

**Present Address**
†Department of Physics and Center for Advanced Nanoscience, University of California, San Diego, La Jolla, California 92093, USA.

ABSTRACT.

We study the electrical transport of vertically-stacked Josephson tunnel junctions using GdBa$_2$Cu$_3$O$_{7-\delta}$ electrodes and a BaTiO$_3$ barrier with thicknesses between 1 nm and 3 nm. The junctions with an area of 20 μm x 20 μm were fabricated combining optical lithography and ion etching using GdBa$_2$Cu$_3$O$_{7-\delta}$ (16 nm) / BaTiO$_3$ (1 - 3 nm) / GdBa$_2$Cu$_3$O$_{7-\delta}$ (16 nm) trilayers growth by sputtering on (100) SrTiO$_3$. Current-voltage measurements at low temperatures show a Josephson coupling for junctions with BaTiO$_3$ barriers of 1 nm and 2 nm. Reducing the barrier thickness bellow a critical thickness seems to suppress the ferroelectric nature of the BaTiO$_3$. The Josephson coupling temperature is strongly reduced for increasing barrier thicknesses, which may be related to the suppression of the superconducting critical temperature in the bottom GdBa$_2$Cu$_3$O$_{7-\delta}$ due to stress. The Josephson energies at 12 K are of ≈ 1.5 mV and ≈ 7.5 mV for BaTiO$_3$ barriers of 1 nm and 2 nm. Fraunhofer patterns are consistent with fluctuations in the critical current due to structural inhomogeneities in the barriers. Our results are promising for the development of Josephson junctions using high-$T_c$ electrodes with energy gaps much higher than those usually present in conventional low-temperature superconductors.

## 1. Introduction

There has been continuous progress in superconductor electronics fabrication towards incorporating materials allowing faster operation speeds, reducing the influence of thermal noise and reducing the minimum size of circuit features [1,2]. Josephson junctions (JJs) based on high-transition-temperature superconductors (HTS) are of technological relevance for many applications going from high-performance computing to high-frequency sensors. The main advantage of HTS over conventional low-temperature superconductors (LTS) is related to the large superconducting gap and high critical temperature ($T_c$), which traduces in devices with high-frequency operation rates and low thermal noise ($I_N$) [3]. Most research in HTS JJs is related to planar arrays of junctions produced by mechanical break [4] and grain boundaries [5,6,7]. Reports on vertically-stacked JJ using HTS are scarcer in the literature [8,9,10,11]. By contrasting planar with vertically-stacked JJs, the latter allow to include barriers with different electrical and magnetic properties [12,13], and to tune critical currents changing the barrier thickness in atomic scales [14]. However, these advantages are conditioned by the 3D growth mechanism usually observed in HTS thin films [15] and by the limitations to fabricate barriers with well-defined nanoscale interfaces [16,17].

Recently, we reported Josephson coupling in superconductor–insulator–superconductor (SIS) fabricated with GdBa$_2$Cu$_3$O$_7$ (GBCO) electrodes and a SrTiO$_3$ barrier [11]. The junctions display high characteristic voltages $V_C = I_c R_n$ (with $I_c$ the critical current and $R_n$, the resistance at the normal state) above those typically observed in those fabricated with LTS [18] and MgB$_2$ [19]. We have found that changing the barrier thickness affects not only the Josephson junction energy, through changes in both the $I_c$ and $R_n$, but also the Josephson temperature. Because materials with a perovskite structure usually display a broad range of physical properties, it may be interesting to extend the study to other barriers. Indeed, new functionalities may be obtained using ferromagnetic [20,21] or ferroelectric materials [13,22,23,24]. In ferromagnetic JJs, so-called pi junctions, the phase of the junction can be set to either 0 or pi depending on the thickness of the ferromagnetic layer [25].On the other hand, two different effects could be expected for a ferroelectric JJs, which is related to the tuning of the critical current. The first is associated with the influence of the polarization on the superconducting properties of the electrodes [23,26]. The second is related to the tuning of the barrier thickness by polarization [27].

In this work, we investigated Josephson coupling in tunnel junctions fabricated using GdBa$_2$Cu3O$_{7-\delta}$ (GBCO) as electrodes and BaTiO$_3$ (BTO) as an insulator barrier. The motivation of this work was studying the influence of a potential ferroelectric barrier on the characteristic current-voltage (IV) curves of JJs at low temperatures. GBCO is a superconducting material with critical temperature $T_c \approx 93$ K. BTO is a ferroelectric material with a Curie temperature of $\approx 390$ K in bulk. For thin films, the transition temperature suppresses as thickness decreases and vanishes at approximately 2 nm [28,29]. GBCO has an orthorhombic structure with lattice parameters of $a = 0.383$ nm, $b = 0.389$ and $c = 1.17$ nm. BTO is a cubic perovskite with $a = 0.399$ nm. Based on our previous work of GBCO/SrTiO$_3$/GBCO junctions [11], the GBCO thickness of the 16 nm was fixed by considering a trade-off between $T_c$ and the presence of smooth surfaces [15]. The BTO thickness was varied between 1 nm and 3 nm. We analyze the junctions by performing IV curves at low temperatures. The presence of Josephson coupling was confirmed from the response of the critical current $I_c$ as a function of the magnetic field (Fraunhofer patterns). The results are discussed considering inhomogeneous barriers produced by thickness fluctuations and interface disorder.

## 2. Experimental

GBCO/BTO/GBCO trilayers were grown on (100) SrTiO$_3$ by sputtering as described in detail elsewhere [15,30,31]. The tunnel junctions were designed using 16 nm thick GBCO electrodes and a BTO barrier with a thickness ($d_{BTO}$) of 1 nm, 2 nm, 3 nm, and 4 nm. During the deposition, the substrate was fixed to the sample holder using silver paint and kept at 730°C in an Ar (90%) / O$_2$ (10%) mixture at a pressure of 400 mTorr. The GBCO and BTO layers were grown using 25 W by DC and RF sources, respectively. After deposition, the temperature decreases in two steps. First, the sample holder is cold down to 500°C, and the O$_2$ pressure increases to 100 Torr. Second, the sample is cold- down to room temperature at a rate of 1.5° C/min. A 2 nm thick STO buffer layer was introduced to reduce the formation of 3D defects in the bottom electrode [15]. Wherever used, the notations [G-$d_{BTO}$-G] indicate a GBCO bottom and top electrodes and a BTO barrier with a thickness $d$ (nm).

X-ray diffraction (XRD) data were obtained using a Panalytical Empyrean diffractometer operated at 40 kV and 30 mA with the Cu$_{K\alpha}$ radiation. The structural analysis was performed based on Θ-2Θ scans with an angular resolution of 0.02°. The thicknesses of the GBCO and BTO layers were estimated from low angle X-ray reflectivity (XRR) measurements (not shown). Atomic force microscopy (AFM) images were obtained in tapping mode with a Dimension 3100 ©Brucker microscope. RMS values correspond to the root mean square average of height deviation taken from the mean image data plane.

Tunnel junctions with an area of 400 μm$^2$ were fabricated according to the steps previously described in reference [11]. Figure 1$a$ shows a schematic diagram of a [G-$d_{BTO}$-G] junction. Before starting with the micro-fabrication process, the sample is covered with 60 nm of silver by sputtering (to avoid surface damage in the top electrode). The fabrication process is as follows: 1-2) using photoresist positive and ion milling we create a bridge with a length of 2 mm and a width of ≈ 70 μm (removing all the sample) and square pillars (20 μm x 20 μm) on the top of the bridge (removing only the top electrode), respectively; 3) positive photoresist is used to create s square pillar of 10 μm x 10 μm on the center of the pillars generated at the first step; 3) the junction is covered with ≈ 100 nm thick SiO$_2$ film by RF sputtering; 4) the area of 10 μm x 10 μm on the top of the junction is open by lift-off process removing the remaining photoresist using acetone; and, 5) a path of silver is deposited by sputtering on the SiO$_2$ capping layer (including the open area of 10 μm x 10 μm) to facilitate electrical connections. Figure 1$b$ shows a picture with the configuration used to make electrical contacts in which the path of silver crosses the junction. Figures 1*cd* show an AFM image and a height profile of the square pillar developed in the steps 1 and 2 mentioned above.

The process to obtain the characteristic current-voltage (IV) curves is similar to the described in reference [11]. The measurements were obtained using the standard four-point geometry. Each point in the curves is the average of 50 measures. A copper coil is used to perform IV curves as a function of the magnetic field. We define the Josephson coupling temperature ($T_J$) as the temperature where the Josephson Effect disappears.

### 3. Results and discussion

Figure 2 shows the XRD patterns for a pure GBCO film, and [G-$d_{BTO}$-G] trilayers with $d$ = 1 nm, 2 nm, and 3 nm. The XRD patterns of the single GBCO layer display the (00l) reflections, indicating epitaxial growth with *c*-axis orientation. The rocking curve for the (005) GBCO line exhibit a full width at half maximum values of 0.35(5)° [15]. The trilayers with $d$ = 2 nm and 3 nm show the reflection (002) of the BTO layer at $2\theta \approx 43.1°$ and 43.7°. The out-plane lattice parameters are $a$ = 0.420 nm ($d$ = 2 nm) and 0.414 nm ($d$ = 3 nm), indicating that the BTO layer is strained (compressed in the plane). The (002) reflection of the BTO is not distinguishable from the background for $d$ = 1 nm, which can be related to the influence of the size in the peak width.

Figure 3 shows the evolution of the surface topology by adding the successive layers measured by AFM in [G-2-G]. The bottom GBCO electrode displays smooth surfaces with a roughness average (Ra) of around 0.7 nm (expressed as root mean square (RMS)) [11]. Typically, GBCO thin films display RMS $\approx$ 0.5 nm. The roughness in the films increases adding the BTO and GBCO capping layers. The GBCO / BTO bilayers show a surface topology with uniform terraces and some 3D defects mainly originated in defects coming from the bottom electrode [31,32]. The borders of 3D defect usually display higher electrical conductivity than terraces increasing the inhomogeneity in the properties of barrier [30,31]. The electrical conductivity across the barrier decreases as the thickness increases. IV curves performed at room temperature show that tunneling and oxygen vacancy migration (OVM) contribute to the conductivity in BTO barriers [32]. The latter suppresses as the temperature reduces [33]. It should be noted that, although the roughness in the GBCO capping layer is irrelevant for the practical ends of the joint, we have found that the inhomogeneous thickness makes more complicated its removal by ion milling during the fabrication of the junction (step 2 described above).

Figure 4 shows the IV curves of JJ with ultra-thin BTO barriers. The JJs with $d$ = 1 nm and 2 nm show Josephson coupling with $T_J$ of approximately 77 K and 41 K, respectively. The suppression of $T_J$ may be associated with both a thicker insulator layer reducing the superconducting wave-function overlap and a lower $T_c$ in the bottom GBCO electrode [11,31]. The $R_N$ values (see criterion in Fig. 4a) are of 0.5 Ω ($d = 1$ nm) and 115 Ω ($d = 2$ nm). The IV curves for [G-3-G] show features related to superconductivity for temperatures lower than 40 K. However, as we previously discussed for a 3 nm thick STO barrier [11], no low field interference effect in the critical current was observed for this sample, ruling out the presence of a Josephson coupling between the superconducting electrodes. In addition, the IV curves in [G-3-G] display a hysteretic behavior at high polarizations (see dotted circle). This effect may be related to OVM [33]. The barrier thickness ($d ≤ 2$nm) in which the Josephson coupling appears using BTO is similar to the previously reported for STO [11], consistent with the short coherence length ξ of the GBCO [34]. The IV curves for [G-1-G] and [G-2-G] display hysteresis, which is a distinctive feature of SIS junctions and is usually called underdamped behavior [35]. There are not ferroelectric features such as hysteretic behavior at the normal state or a step due to resistive switching [23,27,33]. It is important to mention that, although the presence of FE may be affected by the structural disorder, the Josephson coupling disappears for barriers thicker than 2 nm being closer to the limit in which the ferroelectricity vanishes [36].

For further characterization of the junctions, we analyze the critical current $I_c$ as a function of the magnetic field (H). For a rectangular junction with uniform current density,

$$I_c (H) = I_0 \left| \frac{\sin(\frac{\pi H}{H_0})}{\frac{\pi H}{H_0}} \right|, \qquad \text{[eq. 1]}$$

here $I_0 = J_t WL$ ($J_t$ the total current density, $W$ and $L$ the lengths of the junction), $H_0$ is the value of the magnetic field corresponding to a flux quantum penetrating into the junction [35]. Figure 5 shows $I_c$ (H) for [G-1-G] for T = 12 K, 25 K and 50 K. Each point corresponds to an IV curve performed at a fixed H. The criteria for $I_c$ and $R_n$ determination are indicated in Fig. 4a. The curves display the expected modulation for Josephson coupling with minima spacing each $\Delta H ≈ 30$ Oe (see dotted lines in Fig. 5). Similar patterns were obtained for STO barriers [11]. Using equation 1, $\Delta H$ takes place at $\phi_0 = \Delta H \cdot A_{effective} = 30 \cdot L \cdot d$ with $L = 20$ μm for our junctions and $\phi_0 = 2.07 \cdot 10^{-7}$ G·cm⁻², and corresponds to $d ≈ 35$ nm. The latter aggress with the total thickness of the junctions (electrodes with thicknesses of ≈16 nm and a BTO barrier of ≈1-2 nm), which is much smaller than the penetration depth λ in GBCO (λ ≈ 120 nm [34]).

Although the magnetic field response of the junctions displays a qualitative agreement with the predicted by equation 1, a distinctive feature is a residual current at the minima. The origin of this anomaly in the patterns may be related to thickness fluctuations in the BTO layer and inhomogeneous current distribution in the junctions (see Figs 3ab) [11,37,38]. This assumption agrees with conductivity maps obtained at room temperature in GBCO / BTO bilayers indicates that there are fluctuations mainly originated by the presence of topological defects in the bottom GBCO electrode [31]. For inhomogeneous barriers with residual currents at the minima where the thickness fluctuations are small in comparison with the thickness of the barrier, the Fraunhofer pattern has been described by

$$I_c(H) = \sqrt{\left(I_0^2 - \frac{\overline{I_1^2}}{\pi N}\right)\left(\frac{sin\frac{\pi H}{H_0}}{\frac{\pi H}{H_0}}\right)^2 + \frac{\overline{I_1^2}}{\pi N}}, \qquad \text{[eq. 2]}$$

where $\overline{I_1^2}$ is the mean-square of the current fluctuations across the barrier and N is a factor that represent the thickness fluctuation (N > 1 for small fluctuations) [37]. The fits need to considers a factor $\gamma^2 = \left(\frac{\overline{I_1^2}}{I_0^2}\right)\left(\frac{1}{\pi N}\right)$. The data for [G-1-G] at 12 K, 25 K and 50 K reproduce with $\gamma \approx 0.13$, indicating that the mechanism is the same for the three temperatures (see straight lines in Fig. 5).

Figure 6 shows the temperature dependence of $V_c$ for [G-1-G] and [G-2-G]. For comparison, data of reference [11] for HTS and STO barriers with similar thicknesses are included. For SIS with identical superconductors, the theoretical temperature dependence of $V_c$ is limited by the superconducting gap as

$$V_c = I_c R_n = \frac{\pi \Delta(T)}{2e} tanh \frac{\Delta(T)}{2kT}, \qquad \text{[eq. 3]}$$

where $\Delta(T) = \Delta(0) tanh\left\{1.82\left[1.018\left(\frac{T_c}{T} - 1\right)\right]^{0.51}\right\}$ [39] and $\Delta(0) = 1.76kT_c$. Our analysis does not consider non-monotonic effects in the $I_c(T)$ curves usually associated with d-wave superconductors [40]. The Josephson energies at 12 K are of $\approx 1.5$ mV and $\approx 7.5$ mV for BTO barriers with a thickness of 1 nm and 2 nm. These values are strongly affected by the large change in the $R_N$ values of the JJs. As we mentioned above, the $R_N$ value decrease's from 115 $\Omega$ for $d = 2$ nm to 0.5 $\Omega$ for $d = 1$. The low $R_N$ for $d = 1$ infers that the electrode is not fully covered, suggesting the coexistence of SIS and superconductor–normal –superconductor regions (SNS). The $V_c(T)$ dependences decrease systematically to reach zero at $T_J$ of $\approx 77$ K and $\approx 41$ K for $d = 1$ and $d = 2$, respectively. Tunnel junctions with BTO and STO display the same features in $V_c$ and $T_J$. The devices with 1 nm thick barriers exhibit more substantial discrepancies than the samples with thicker barriers. This effect may be related to the higher influence of fluctuations in the barrier thickness and interface effects on the properties of ultra-thin barriers (coexistence of conducting and insulator regions). Moreover, as in STO layers, even for a low $T_J$, the best performance is found for the thicker barrier where low density of pinholes and conducting regions are expected [11,31].

The large Josephson energy for the samples reported here and in reference [11] is evidenced in the low thermal noise of the IV curves. The value of $V_c$ (12 K) $\approx 7.5$ mV is larger than that generally observed in NbN [18] and MgB2 [19,41] tunnel junctions. Among nitrides, NbN displays the maximum $\Delta(0)$ with a theoretical limit of $V_c \approx 4$ mV. On the other hand, a value of $\approx 2$ mV has been experimentally reported for MgB$_2$ [19,41]. Moreover, the Josephson energies using BTO and STO barriers result in higher than those reported in planar junctions [42,43]. Even considering that $T_J$ in our junctions reduces as barrier thickness increases, the high $V_c$ values offer several advantages for operation at low temperatures. Indeed, the quantum effects tend to wash out as a consequence of thermal noise when the operation temperature is increased, indicating that low operating temperatures are desired even for JJ with HTS. As we mentioned above, a high Josephson energy is essential for the development of high-quality JJ for two factors, reduced influence of thermal noise and a higher operating frequency.

Our results in Josephson junctions fabricated using HTS and insulator barriers such as BTO and STO have demonstrated similar behavior [11]. We found that at low temperatures, the JJs display higher $V_C$ as the barrier thickness increases. The Josephson coupling usually vanishes as the barrier thickness increases to 2 nm, which is consistent with the short coherence length $\xi$ of the HTS. The high $V_c$ values displayed by vertically-stacked JJ using HTS is promising for applications in electronic systems that require a high-frequency operation. As a side note, two points should be considered towards improving the performance of our results. First, higher $I_c$ values should be obtained for 2 nm thick insulator barriers increasing $T_c$ in the bottom electrode, this should increase the Josephson energy of the Junctions further improving its operating frequency and reducing the influence of thermal noise. Second, it is essential to note that the properties in ultra-thin films based on perovskites usually are affected by interface disorder. This fact implies a technological challenge associated with the optimization of the interfaces to conserve bulk properties or to generate new ones in the nanometer scale.

## 4. Conclusions

We characterized the Josephson coupling in HTS junctions using ultra-thin BaTiO3, which is an essential step towards the integration of perovskite barriers in superconducting JJ. The results show that tunnel junctions with a barrier thickness of 1 nm and 2 nm display the expected behavior for the Josephson Effect. The critical barrier thickness for ferroelectric effects in these systems is usually higher than 2 nm. The Josephson energies at 12 K are of »1.5 mV and » 7.5 mV for BaTiO3 barriers of 1 nm and 2 nm. Fraunhofer patterns are consistent with fluctuations in the critical current due to inhomogeneities in the insulator barrier. The present experiments show Josephson coupling with energies above the theoretical limits for LTS and similar to the expectations for ideal junctions with $MgB_2$. When searching for new functionalities using perovskite compounds with different physical properties (i.e. ferroelectric or ferromagnetic barriers), special care should be taken oriented to optimize their properties at the nanoscale.

**Acknowledgements**

This work was partially supported by the ANPCYT PICT 2015-2171 and 2018-03126. Jeehoon Kim was supported by the National Research Foundation of Korea (NRF) and the Korea government (MSIT) (grants 2016K1A4A4A01922136, 2018R1A5A6075964, 2019R1A2C2090356). MS and NH are members of the Instituto de Nanociencia y Nanotecnología CNEA-CONICET (Argentina).

Figure 1. *a)* Schematic diagram of a [G-$d_{BTO}$-G] junction. *b)* Image of the tunnel junction and the silver path used to facilitate the electrical contacts in the top GBCO electrode. *c)* AFM image and a height profile of the square pillar after remove the top GBCO electrode.

Figure 2. X-ray diffraction patterns (logarithmic intensity scale) of a 16 nm thick GBCO films and [G-$d_{BTO}$-G] trilayers with $d = 1$ nm, 2 nm, and 3 nm. The measurements were performed at room temperature.

Figure 3. 10 x 10 $\mu m^2$ topographical images with the surface evolution including a 16 nm thick GBCO thin with the successive addition of a 2 nm thick BTO insulator barrier and a 16 nm GBCO capping layer. Surface roughness height profiles are included.

Figure 4. Characteristic IV curves at different temperatures for a) [G-1-G]; b) [G-2-G]; and, c) [G-3-G]. Measurements are performed applying current and measuring voltage. Josephson coupling is observed for JJ with BTO barriers of 1 nm and 2 nm. The criteria for the determination of $I_c$ and $R_n$ are indicated in *a)*. The inset panel a) corresponds to a zoom of the IV curve at 76 K. The arrows in the hysteretic regions of the curves indicate the direction in the measurements.

Figure 5. $I_c$ modulation by an external magnetic field for [G-1-G] at *a)* 12 K: *b)* 25 K: and, *c)* 50 K. The dotted and straight lines correspond to the fit for an ideal Josephson junction using equation 1 and equation 2, respectively.

Figure 6. Comparison of $V_c$ measured in [G-1-G] and [G-2-G] with the theoretical expectations according to the equation 2. The dashed lines correspond to the expected values considering $T_J$ of 42 K and 76 K. Data for JJ with STO barriers of 1 nm and 2 nm from reference [11] are included.

Figure 1.

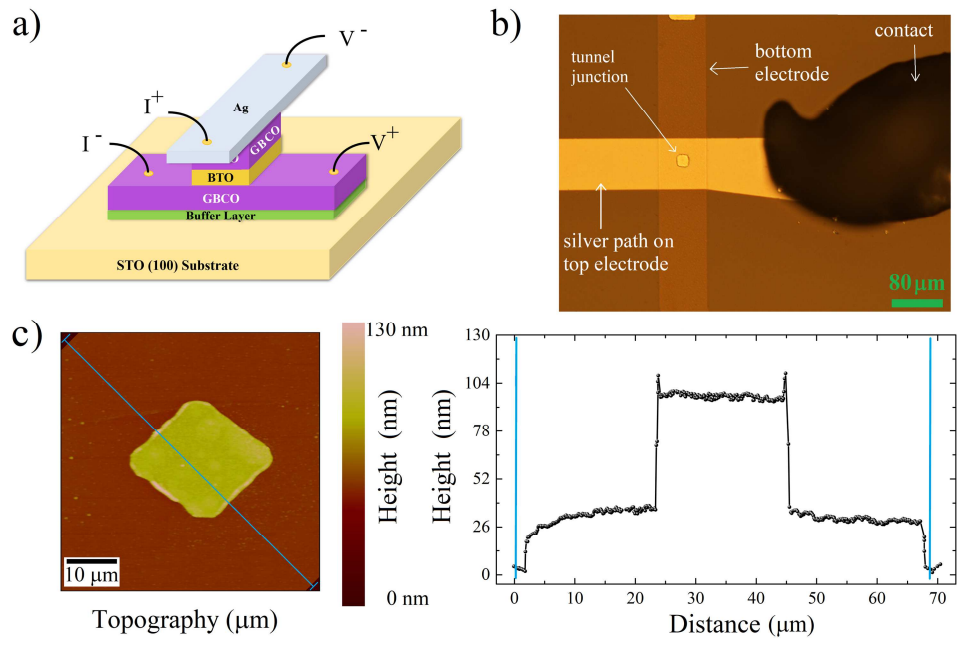

Figure 2.

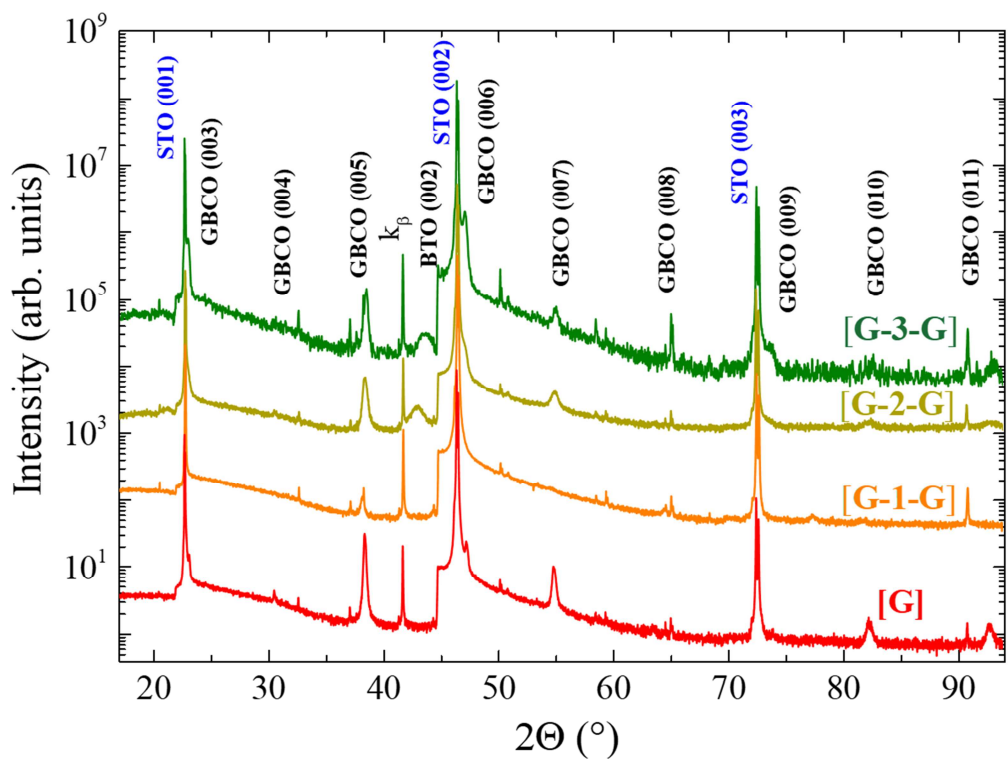

Figure 3.

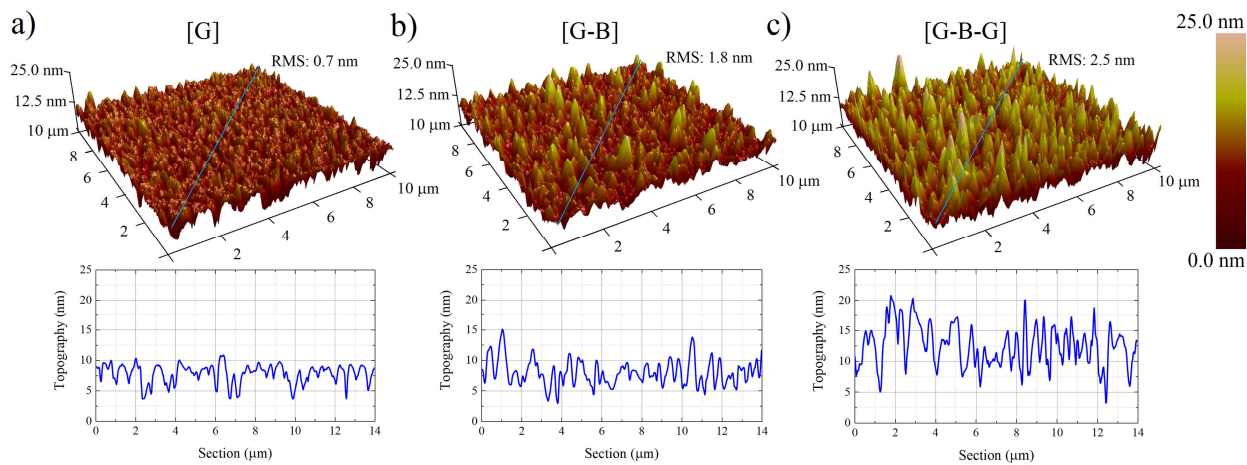

Figure 4.

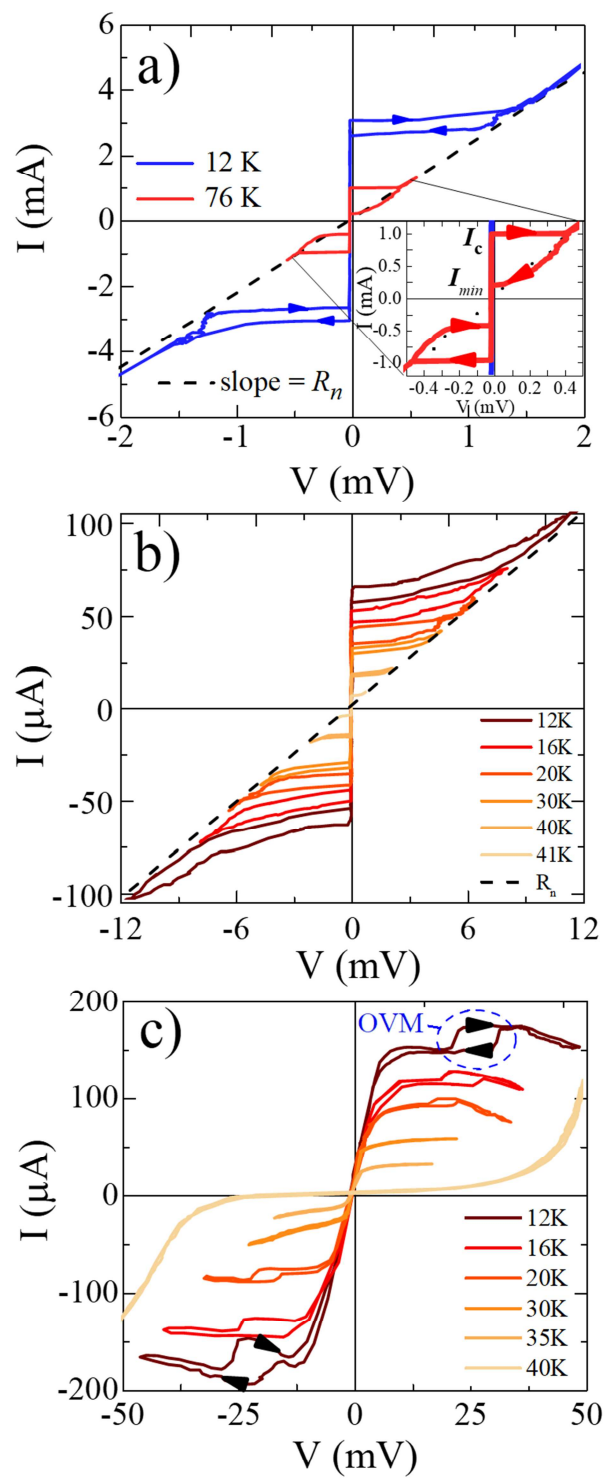

Figure 5.

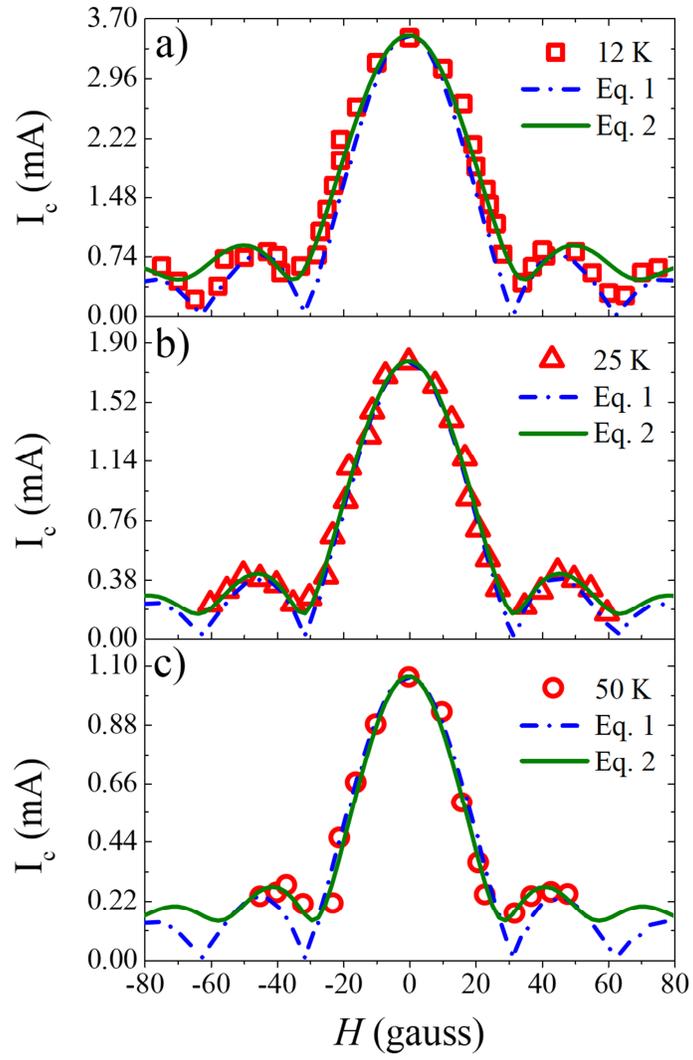

Figure 6.

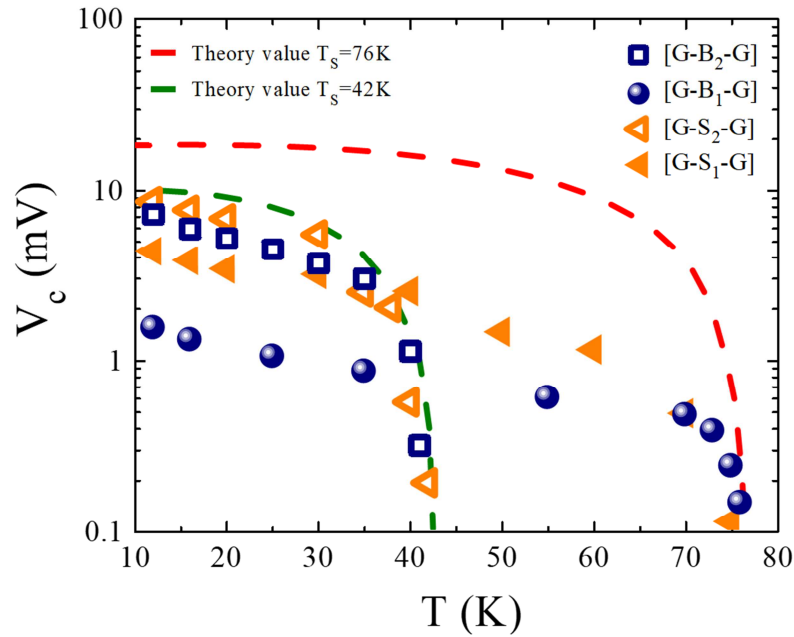